\documentclass[12pt]{article}%
\usepackage{amsmath}
\usepackage{amsfonts}
\usepackage{amssymb}
\usepackage{graphicx}%
\newtheorem{theorem}{Theorem}

\newtheorem{example}[theorem]{Example}

\begin{document}

\title{On Beltrami fields with nonconstant proportionality factor on the plane}
\author{Vladislav V. Kravchenko$^{\text{1}}$, H\'{e}ctor Oviedo$^{\text{2}}$\\$^{\text{1}}${\small Department of Mathematics, CINVESTAV del IPN, Unidad
Queretaro, }\\{\small Libramiento Norponiente No. 2000, Fracc. Real de Juriquilla,
Queretaro, }\\{\small Qro. C.P. 76230 MEXICO e-mail:
vkravchenko@qro.cinvestav.mx\thanks{Research was supported by CONACYT, Mexico
via the research project 50424.}}\\$^{\text{2}}${\small SEPI, ESIME Zacatenco, Instituto Polit\'{e}cnico
Nacional, Av. IPN S/N, }\\{\small C.P. 07738, D.F. MEXICO\thanks{During the preparation of this work the
second-named author was supported by CONACYT on a postdoctoral stay at the
Department of Mathematics, CINVESTAV del IPN, Unidad Quer\'{e}taro}}}
\maketitle

\begin{abstract}
We consider the equation
\begin{equation}
\operatorname{rot}\overrightarrow{B}+\alpha\overrightarrow{B}=0 \label{1}%
\end{equation}
on the plane with $\alpha$ being a real-valued function and show that it can
be reduced to a Vekua equation of a special form. In the case when $\alpha$
depends on one Cartesian variable a complete system of exact solutions of the
Vekua equation and hence of equation (1) is constructed based on L. Bers'
theory of pseudoanalytic formal powers.

\end{abstract}

\section{Introduction}

Solutions of the equation
\begin{equation}
\operatorname{rot}\overrightarrow{B}+\alpha\overrightarrow{B}%
=0\label{Beltrami}%
\end{equation}
where $\alpha$ is a scalar function of space coordinates are known as Beltrami
fields and are of fundamental importance in different branches of modern
physics (see, e.g., \cite{Zag}, \cite{Lak}, \cite{Feng}, \cite{Yoshida},
\cite{Stratis}, \cite{Kaiser}, \cite{Gonzalez}, \cite{KrBeltrami}). For
simplicity, in this work we consider the real-valued proportionality factor
$\alpha$ and real-valued solutions of (\ref{Beltrami}), though the presented
approach is applicable in a complex-valued situation as well with a
considerable complication of mathematical techniques involved (instead of
complex Vekua equations their bicomplex generalizations should be considered
\cite{KrAntonio}, \cite{KrJPhys06}). We consider equation (\ref{Beltrami}) on
a plane of the variables $x$ and $y$, that is $\alpha$ and $\overrightarrow
{B}$ are functions of two Cartesian variables only. In this case as we show in
section \ref{SectReduction} equation (\ref{Beltrami}) reduces to the equation
\begin{equation}
\operatorname*{div}\left(  \frac{1}{\alpha}\nabla u\right)  +\alpha
u=0.\label{a-1aeq}%
\end{equation}
This second-order equation can be reduced (see \cite{KrJPhys06}) to a
corresponding Vekua equation (describing generalized analytic functions) of a
special form. This reduction under quite general conditions allows us to
construct a complete system of exact solutions of (\ref{a-1aeq}) explicitly
(see \cite{Krpseudoan} and \cite{KrRecentDevelopments}). For the reduction of
(\ref{a-1aeq}) to a Vekua equation it is sufficient to find a particular
solution of (\ref{a-1aeq}). In the present work (section
\ref{Sect_alpha_one_variable}) we show that in a very important for
applications case of $\alpha$ being a function of one Cartesian variable a
particular solution of (\ref{a-1aeq}) is always available in a simple explicit
form. This situation corresponds to models describing waves propagating in
stratified media (see, e.g., \cite{KO2003}). As a result in this case we are
able to construct a complete system of solutions explicitly which for many
purposes means a general solution. We give an example of such construction.

We show in this work that when $\alpha=\alpha(y)$ (of course in a similar way
the case $\alpha=\alpha(x)$ can be considered) equation (\ref{a-1aeq}) and
hence equation (\ref{Beltrami}) reduce to the Vekua equation of the following
form%
\begin{equation}
\partial_{\overline{z}}W(x,y)=\frac{if^{\prime}(y)}{2f(y)}\overline
{W}(x,y)\label{Vekuaintro}%
\end{equation}
where
\[
f=\frac{c_{1}}{\sqrt{\alpha}}\sin\mathcal{A}+\frac{c_{2}}{\sqrt{\alpha}}%
\cos\mathcal{A};
\]
$\mathcal{A}$ is an antiderivative of $\alpha$ with respect to $y$, $c_{1}$
and $c_{2}$ are arbitrary real constants, $z=x+iy$ and $\partial_{\overline
{z}}=\frac{1}{2}(\partial_{x}+i\partial_{y})$. A complete (in a compact
uniform convergence topology) system of exact solutions to (\ref{Vekuaintro})
can be constructed explicitly. The system represents a set of formal powers
\cite{Berskniga}, \cite{Courant} which generalize the usual analytic complex
powers $(z-z_{0})^{n},$ $n=0,1,2,\ldots$ and in a sense give us a general
solution of (\ref{Vekuaintro}). Thus, in the case when $\alpha$ is a function
of one Cartesian variable the Vekua equation equivalent to (\ref{Beltrami}) in
a two-dimensional situation can be solved and a complete system of solutions
of (\ref{Beltrami}) is obtained.

\section{Preliminaries\label{SectPreliminary}}

We need the following definition. Consider the equation
\begin{equation}
\partial_{\overline{z}}\varphi=\Phi\label{dzphi}%
\end{equation}
on a whole complex plane or on a convex domain, where $\Phi=\Phi_{1}+i\Phi
_{2}$ is a given complex valued function such that its real part $\Phi_{1}$
and imaginary part $\Phi_{2}$ satisfy the equation
\begin{equation}
\partial_{y}\Phi_{1}-\partial_{x}\Phi_{2}=0\label{casirot}%
\end{equation}
then there exist real valued solutions of (\ref{dzphi}) which can be easily
constructed in the following way%
\begin{equation}
\varphi(x,y)=2\left(  \int_{x_{0}}^{x}\Phi_{1}(\eta,y)d\eta+\int_{y_{0}}%
^{y}\Phi_{2}(x_{0},\xi)d\xi\right)  +c\label{Antigr}%
\end{equation}
where $(x_{0},y_{0})$ is an arbitrary fixed point in the domain of interest
and $c$ is an arbitrary real constant.

By $\overline{A}$ we denote the integral operator in (\ref{Antigr}):%
\[
\overline{A}[\Phi](x,y)=2\left(  \int_{x_{0}}^{x}\Phi_{1}(\eta,y)d\eta
+\int_{y_{0}}^{y}\Phi_{2}(x_{0},\xi)d\xi\right)  +c.
\]
Note that formula (\ref{Antigr}) can be extended to any simply connected
domain by considering the integral along an arbitrary rectifiable curve
$\Gamma$ leading from $(x_{0},y_{0})$ to $(x,y)$%
\[
\varphi(x,y)=2\left(  \int_{\Gamma}\Phi_{1}dx+\Phi_{2}dy\right)  +c.
\]
Thus if $\Phi$ satisfies (\ref{casirot}), there exists a family of real valued
functions $\varphi$ such that $\partial_{\overline{z}}\varphi=\Phi$, given by
the formula $\varphi=\overline{A}[\Phi]$.

Let $f$ denote a given positive twice continuously differentiable function
defined on a domain $\Omega\subset\mathbb{C}$. Consider the following Vekua
equation%
\begin{equation}
W_{\overline{z}}=\frac{f_{\overline{z}}}{f}\overline{W}\qquad\text{in }%
\Omega\label{Vekuamain}%
\end{equation}
where the subindex $\overline{z}$ means the application of the operator
$\partial_{\overline{z}}$, $W$ is a complex-valued function and $\overline{W}$
is its complex conjugate function. As was shown in \cite{KrBers},
\cite{Krpseudoan}, \cite{KrJPhys06}, \cite{KrRecentDevelopments},
\cite{KrOviedo06} equation (\ref{Vekuamain}) is closely related to the
second-order equation of the form
\begin{equation}
(\operatorname{div}p\operatorname{grad}+q)u=0\text{\qquad in }\Omega
\label{maineq}%
\end{equation}
where $p$ and $q$ are real-valued functions. In particular the following
statements are valid.

\begin{theorem}
\cite{KrJPhys06} Let $u_{0}$ be a positive solution of (\ref{maineq}). Assume
that $f=\sqrt{p}u_{0}$ and $W$ is any solution of (\ref{Vekuamain}). Then
$u=\frac{1}{\sqrt{p}}\operatorname*{Re}W$ is a solution of (\ref{maineq}) and
$v=\sqrt{p}\operatorname*{Im}W$ is a solution of
\begin{equation}
(\operatorname{div}\frac{1}{p}\operatorname{grad}+q_{1})v=0\text{\qquad in
}\Omega\label{assocmaineq}%
\end{equation}
where
\begin{equation}
q_{1}=-\frac{1}{p}\left(  \frac{q}{p}+2\left\langle \frac{\nabla p}{p}%
,\frac{\nabla u_{0}}{u_{0}}\right\rangle +2\left(  \frac{\nabla u_{0}}{u_{0}%
}\right)  ^{2}\right)  . \label{q1}%
\end{equation}

\end{theorem}

\begin{theorem}
\cite{KrJPhys06} Let $\Omega$ be a simply connected domain, $u_{0}$ be a
positive solution of (\ref{maineq}) and $f=p^{1/2}u_{0}$. Assume that $u$ is a
solution of (\ref{maineq}). Then a solution $v$ of (\ref{assocmaineq}) with
$q_{1}$ defined by (\ref{q1}) such that $W=p^{1/2}u+ip^{-1/2}v$ is a solution
of (\ref{Vekuamain}) is constructed according to the formula%
\begin{equation}
v=u_{0}^{-1}\overline{A}(ipu_{0}^{2}\partial_{\overline{z}}(u_{0}%
^{-1}u))\label{darboux1}%
\end{equation}
and vice versa, let $v$ be a solution of (\ref{assocmaineq}), then the
corresponding solution $u$ of (\ref{maineq}) such that $W=p^{1/2}u+ip^{-1/2}v$
is a solution of (\ref{Vekuamain}), is constructed according to the formula%
\begin{equation}
u=-u_{0}\overline{A}(ip^{-1}u_{0}^{-2}\partial_{\overline{z}}(u_{0}%
v)).\label{darboux2}%
\end{equation}

\end{theorem}

Thus the relation between (\ref{Vekuamain}) and (\ref{maineq}) is very similar
to that between the Cauchy-Riemann system and the Laplace equation. Moreover,
choosing $p\equiv1$, $q\equiv0$ and $u_{0}\equiv1$ we obtain that
(\ref{darboux1}) and (\ref{darboux2}) become the well known formulas from the
classical complex analysis for constructing conjugate harmonic functions.

For a Vekua equation of the form%
\[
W_{\overline{z}}=aW+b\overline{W}%
\]
where $a$ and $\ b$ are arbitrary complex-valued functions from an appropriate
function space \cite{Vekua} a well developed theory of Taylor and Laurent
series in formal powers was created (see \cite{Berskniga},
\cite{BersFormalPowers}) containing among others the expansion and the Runge
theorems as well as more precise convergence results (see, e.g., \cite{Menke})
and a general simple algorithm \cite{KrRecentDevelopments} for explicit
construction of formal powers for the Vekua equation of the form
(\ref{Vekuamain}).

\section{Reduction of (\ref{Beltrami}) to a Vekua
equation\label{SectReduction}}

We consider equation (\ref{Beltrami}) where both $\alpha$ and $\overrightarrow
{B}$ are supposed to be dependent on two Cartesian variables $x$ and $y$. Then
equation (\ref{Beltrami}) can be written as the following system%
\begin{equation}
\partial_{y}B_{3}+\alpha B_{1}=0 \label{Beltrami1}%
\end{equation}%
\begin{equation}
-\partial_{x}B_{3}+\alpha B_{2}=0 \label{Beltrami2}%
\end{equation}%
\[
\partial_{x}B_{2}-\partial_{y}B_{1}+\alpha B_{3}=0.
\]
Solving this system for $B_{3}$ leads to the equation%
\begin{equation}
\Delta B_{3}-\left\langle \frac{\nabla\alpha}{\alpha},\nabla B_{3}%
\right\rangle +\alpha^{2}B_{3}=0 \label{vsp11}%
\end{equation}
where $\left\langle \cdot,\cdot\right\rangle $ denotes the usual scalar
product of two vectors.

Note that \
\[
\alpha\operatorname*{div}\left(  \frac{1}{\alpha}\nabla B_{3}\right)  =\Delta
B_{3}-\left\langle \frac{\nabla\alpha}{\alpha},\nabla B_{3}\right\rangle
\]
and hence (\ref{vsp11}) can be rewritten as follows
\begin{equation}
\operatorname*{div}\left(  \frac{1}{\alpha}\nabla B_{3}\right)  +\alpha
B_{3}=0. \label{maineqalpha}%
\end{equation}
Thus equation (\ref{Beltrami}) reduces to an equation of the form
(\ref{maineq}) with $p=1/\alpha$ and $q=\alpha$.

Let us notice that (see, e.g., \cite{KrJPhys06})%
\[
\operatorname*{div}\frac{1}{\alpha}\nabla+\alpha=\frac{1}{\sqrt{\alpha}%
}\left(  \Delta-r\right)  \frac{1}{\sqrt{\alpha}}%
\]
where
\begin{equation}
r=-\frac{1}{2}\frac{\Delta\alpha}{\alpha}+\frac{3}{4}\left(  \frac
{\nabla\alpha}{\alpha}\right)  ^{2}-\alpha^{2}. \label{r}%
\end{equation}
That is $B_{3}$ is a solution of (\ref{maineqalpha}) iff the function
$f=B_{3}/\sqrt{\alpha}$ is a solution of the stationary Schr\"{o}dinger
equation%
\begin{equation}
\left(  -\Delta+r\right)  f=0 \label{Schrodalpha}%
\end{equation}
with $r$ defined by (\ref{r}). As was explained in section
\ref{SectPreliminary}, given its particular solution this equation reduces to
the Vekua equation (\ref{Vekuamain}). Unfortunately, in general we are not
able to propose a particular solution of (\ref{maineqalpha}). Nevertheless in
an important special case when $\alpha$ depends on one Cartesian variable, a
particular solution of (\ref{maineqalpha}) is always available in explicit
form. We give this result in the next section.

\section{Solution in the case when $\alpha$ is a function of one Cartesian
variable\label{Sect_alpha_one_variable}}

Let us consider equation (\ref{Schrodalpha}) where $\alpha=\alpha(y)$. We
assume that $\alpha$ is a nonvanishing function and look for a solution of the
corresponding ordinary differential equation
\[
\frac{d^{2}f_{0}}{dy^{2}}+\left(  \frac{1}{2}\frac{\alpha^{\prime\prime}%
}{\alpha}-\frac{3}{4}\left(  \frac{\alpha^{\prime}}{\alpha}\right)
^{2}+\alpha^{2}\right)  f_{0}=0.
\]
Its general solution is known (see \cite[2.162 (14)]{Kamke}) and is given by
the expression
\begin{equation}
f_{0}(y)=\frac{c_{1}}{\sqrt{\alpha(y)}}\sin\mathcal{A}(y)+\frac{c_{2}}%
{\sqrt{\alpha(y)}}\cos\mathcal{A}(y) \label{gensolSchr}%
\end{equation}
where $\mathcal{A}$ is an antiderivative of $\alpha$ and $c_{1}$, $c_{2}$ are
arbitrary real constants.

Choosing, e.g., $c_{1}=1$, $c_{2}=0$ and calculating the coefficient $\left(
\partial_{\overline{z}}f_{0}\right)  /f_{0}$ we arrive at the following Vekua
equation which is equivalent to (\ref{Beltrami}) in the case under
consideration (and which is considered in any simply connected domain where
$\sin\mathcal{A}(y)$ does not vanish),
\[
\partial_{\overline{z}}W(x,y)=\frac{i}{2}\left(  \alpha(y)\cot\mathcal{A}%
(y)-\frac{\alpha^{\prime}(y)}{2\alpha(y)}\right)  \overline{W}(x,y).
\]
Note that $F=f_{0}=\frac{\sin\mathcal{A}(y)}{\sqrt{\alpha(y)}}$ and
$G=\frac{i}{f_{0}}=\frac{i\sqrt{\alpha(y)}}{\sin\mathcal{A}(y)}$ represent a
generating pair for this Vekua equation (see \cite{Krpseudoan},
\cite{KrRecentDevelopments}) and hence if $W$ is its solution, the
corresponding pseudoanalytic function of the second kind $\omega=\frac
{1}{f_{0}}\operatorname*{Re}W+if_{0}\operatorname*{Im}W$ satisfies the
equation
\begin{equation}
\omega_{\overline{z}}=\frac{1-f_{0}^{2}}{1+f_{0}^{2}}\overline{\omega
}_{\overline{z}} \label{seckind}%
\end{equation}
which can be written in the form of the following system%
\[
\phi_{x}=\frac{1}{f_{0}^{2}}\psi_{y},\qquad\phi_{y}=-\frac{1}{f_{0}^{2}}%
\psi_{x}%
\]
where $\phi=\operatorname*{Re}\omega$ and $\psi=\operatorname*{Im}\omega$.

For $f_{0}$ being representable in a separable form $f_{0}(x,y)=X(x)Y(y)$ the
formulas for constructing corresponding formal powers explicitly were
presented already by L. Bers and A. Gelbart (see \cite{Berskniga} and
\cite{Courant}). Using them we obtain the following representation for the
formal powers corresponding to (\ref{seckind})%
\begin{align*}
_{\ast}Z^{(n)}(a,z_{0};z) &  =a_{1}%
{\displaystyle\sum\limits_{k=0}^{n}}
\left(
\begin{array}
[c]{c}%
n\\
k
\end{array}
\right)  (x-x_{0})^{(n-k)}i^{k}Y^{k}\\
&  +ia_{2}%
{\displaystyle\sum\limits_{k=0}^{n}}
\left(
\begin{array}
[c]{c}%
n\\
k
\end{array}
\right)  (x-x_{0})^{(n-k)}i^{k}\widetilde{Y}^{k}\text{ \ }%
\end{align*}
(we preserve the notations from \cite{Berskniga}) where $z_{0}=x_{0}+iy_{0}$
is an arbitrary point of the domain of interest, $a$ is an arbitrary complex
number: $a=a_{1}+ia_{2}$, $Y^{k}$ and $\widetilde{Y}^{k}$ are constructed as
follows%
\[
Y^{(0)}(y_{0},y)=\widetilde{Y}^{(0)}(y_{0},y)=1
\]
and for $n=1,2,\ldots$
\[
Y^{(n)}(y_{0},y)=n%
{\displaystyle\int\limits_{y_{0}}^{y}}
Y^{(n-1)}(y_{0},\eta)f_{0}^{2}(\eta)d\eta\text{ \ \ \ \ }n\text{ \ odd}%
\]%
\[
Y^{(n)}(y_{0},y)=n%
{\displaystyle\int\limits_{y_{0}}^{y}}
Y^{(n-1)}(y_{0},\eta)\frac{d\eta}{f_{0}^{2}(\eta)}\text{ \ \ \ \ }n\text{
\ even}%
\]%
\[
\widetilde{Y}^{(n)}(x_{0},x)=n%
{\displaystyle\int\limits_{y_{0}}^{y}}
\widetilde{Y}^{(n-1)}(y_{0},\eta)\frac{d\eta}{f_{0}^{2}(\eta)}\text{
\ \ \ \ }n\text{ \ odd}%
\]%
\[
\widetilde{Y}^{(n)}(x_{0},x)=n%
{\displaystyle\int\limits_{y_{0}}^{y}}
\widetilde{Y}^{(n-1)}(y_{0},\eta)f_{0}^{2}(\eta)d\eta\text{ \ \ \ \ }n\text{
\ even.}%
\]
The system $\left\{  _{\ast}Z^{(n)}(1,z_{0};z),\,_{\ast}Z^{(n)}(i,z_{0}%
;z)\right\}  _{n=0}^{\infty}$ represents a complete (in a compact uniform
convergence topology \cite{BersexpansionMonogenic}) system of solutions of
(\ref{seckind}) that means that any solution $\omega$ of (\ref{seckind}) in a
simply connected domain $\Omega$\ can be represented as a series
\[
\omega(z)=\sum_{n=0}^{\infty}\,_{\ast}Z^{(n)}(a_{n},z_{0};z)=\sum
_{n=0}^{\infty}\left(  a_{n}^{\prime}\,_{\ast}Z^{(n)}(1,z_{0};z)+a_{n}%
^{\prime\prime}\,_{\ast}Z^{(n)}(i,z_{0};z)\right)
\]
where $a_{n}^{\prime}=\operatorname*{Re}a_{n}$, $a_{n}^{\prime\prime
}=\operatorname*{Im}a_{n}$ and the series converges normally (uniformly on any
compact subset of $\Omega$). Consequently the system of functions
\[
\left\{  f_{0}(y)\operatorname*{Re}(_{\ast}Z^{(n)}(1,z_{0};z)),\quad
\,f_{0}(y)\operatorname*{Re}(_{\ast}Z^{(n)}(i,z_{0};z))\right\}
_{n=0}^{\infty}%
\]
represents in the same sense a complete system of solutions of
(\ref{Schrodalpha}) with $r$ defined by (\ref{r}), and
\begin{equation}
\left\{  \sqrt{\alpha(y)}f_{0}(y)\operatorname*{Re}(_{\ast}Z^{(n)}%
(1,z_{0};z)),\quad\sqrt{\alpha(y)}f_{0}(y)\operatorname*{Re}(_{\ast}%
Z^{(n)}(i,z_{0};z))\right\}  _{n=0}^{\infty}\label{complsysmaineqalpha}%
\end{equation}
is a complete system of solutions of (\ref{maineqalpha}). Thus in the case
under consideration any solution $B_{3}$ of (\ref{maineqalpha}) can be
represented in the form
\[
B_{3}(x,y)=\sum_{n=0}^{\infty}(a_{n}\sin\mathcal{A}(y)\operatorname*{Re}%
(_{\ast}Z^{(n)}(1,z_{0};z))+b_{n}\sin\mathcal{A}(y)\operatorname*{Re}(_{\ast
}Z^{(n)}(i,z_{0};z)))
\]
where $a_{n}$ and $b_{n}$ are real constants.

The other two components of the vector $\overrightarrow{B}$ are obtained from
(\ref{Beltrami1}) and (\ref{Beltrami2}):%
\begin{equation}
B_{1}=-\frac{1}{\alpha}\partial_{y}B_{3}\quad\text{and}\quad B_{2}=\frac
{1}{\alpha}\partial_{x}B_{3}\label{B1B2}%
\end{equation}
that gives us a complete system of solutions of (\ref{Beltrami}) in the case
under consideration. On the following example we explain how this procedure
works. 

\begin{example}
\label{ExBeltrami}Let us consider the following relatively simple situation in
which the corresponding integrals are not difficult to evaluate. Let
\begin{equation}
\alpha(y)=\frac{1}{\sqrt{1-y^{2}}} \label{alphaexample}%
\end{equation}
and $\Omega$ be an open unitary disk with a center in the origin. We take in
(\ref{gensolSchr}) $c_{1}=0$ and $c_{2}=1$. Then it is easy to verify that
\[
f_{0}(y)=(1-y^{2})^{\frac{3}{4}}.
\]
The first three formal powers with a centre in the origin can be calculated as
follows%
\[
_{\ast}Z^{(1)}(1,0;z)=x+i[\frac{y(1-y^{2})^{\frac{3}{2}}}{4}+\frac
{3y(1-y^{2})^{\frac{1}{2}}}{8}+\frac{3}{8}\arcsin y],
\]%
\[
_{\ast}Z^{(1)}(i,0;z)=-\frac{y}{(1-y^{2})^{\frac{1}{2}}}+ix,
\]%
\begin{align*}
_{\ast}Z^{(2)}(1,0;z)  &  =x^{2}-\frac{1}{4}y^{2}-\frac{3}{4}\frac{y\arcsin
y}{(1-y^{2})^{\frac{1}{2}}}\\
&  +2ix\left(  \frac{y(1-y^{2})^{\frac{3}{2}}}{4}+\frac{3y(1-y^{2})^{\frac
{1}{2}}}{8}+\frac{3}{8}\arcsin y\right)  ,
\end{align*}%
\[
_{\ast}Z^{(2)}(i,0;z)=-\frac{2xy}{(1-y^{2})^{\frac{1}{2}}}+i\left(
x^{2}-y^{2}-\frac{1}{2}y^{4}\right)  ,
\]%
\begin{align*}
_{\ast}Z^{(3)}(1,0;z)  &  =x^{3}-3x\left(  \frac{1}{4}y^{2}+\frac{3}{4}%
\frac{y\arcsin y}{(1-y^{2})^{\frac{1}{2}}}\right) \\
&  +3ix^{2}\left(  \frac{y(1-y^{2})^{\frac{3}{2}}}{4}+\frac{3y(1-y^{2}%
)^{\frac{1}{2}}}{8}+\frac{3}{8}\arcsin y\right) \\
&  -i(-\frac{3}{24}y(1-y^{2})^{\frac{5}{2}}+\frac{3}{96}y(1-y^{2})^{\frac
{3}{2}}+y(1-y^{2})^{\frac{1}{2}}(\frac{51}{128}-\frac{9}{64}y^{2})\\
&  -\frac{9}{16}(1-y^{2})^{2}\arcsin y+\frac{33}{128}\arcsin y),
\end{align*}%
\[
_{\ast}Z^{(3)}(i,0;z)=-\frac{3x^{2}y}{(1-y^{2})^{\frac{1}{2}}}+\frac{3}%
{4}\frac{y(1+y^{2})}{(1-y^{2})^{\frac{1}{2}}}-\frac{3}{4}\arcsin y+ix\left(
x^{2}-3\left(  y^{2}-\frac{1}{2}y^{4}\right)  \right)  .
\]
Now taking the real parts of these formal powers and multiplying them by the
factor $\sqrt{\alpha}f_{0}$ (see (\ref{complsysmaineqalpha})) we obtain the
first elements of the complete system of solutions of (\ref{maineqalpha}),
that is any solution $B_{3}$ of (\ref{maineqalpha}) in a simply connected
domain can be represented as an infinite linear combination of the functions%
\begin{align*}
&  \{(1-y^{2})^{\frac{1}{2}},\quad x(1-y^{2})^{\frac{1}{2}},\quad
-y,\quad(1-y^{2})^{\frac{1}{2}}\left(  x^{2}-\frac{1}{4}y^{2}-\frac{3}{4}%
\frac{y\arcsin y}{(1-y^{2})^{\frac{1}{2}}}\right)  ,\\
&  -2xy,\quad(1-y^{2})^{\frac{1}{2}}\left(  x^{3}-3x\left(  \frac{1}{4}%
y^{2}+\frac{3}{4}\frac{y\arcsin y}{(1-y^{2})^{\frac{1}{2}}}\right)  \right)
,\\
&  -3x^{2}y+\frac{3}{4}y(1+y^{2})-\frac{3}{4}(1-y^{2})^{\frac{1}{2}}\arcsin
y,...\}
\end{align*}
and the corresponding series converges normally.

From (\ref{B1B2}) it is easy to calculate the corresponding components $B_{1}$
and $B_{2}$ respectively,
\begin{align*}
&  \{y,\quad xy,\quad(1-y^{2})^{\frac{1}{2}},\quad\frac{3}{4}(1-y^{2}%
)^{\frac{1}{2}}\arcsin y+y(x^{2}-\frac{3}{4}y^{2}+\frac{5}{4}),\\
&  2x(1-y^{2})^{\frac{1}{2}},\quad\frac{9}{4}x(1-y^{2})^{\frac{1}{2}}\arcsin
y+y\left(  x^{3}-\frac{9}{4}xy^{2}+\frac{15}{4}x\right)  ,\\
&  -(\frac{9}{4}y^{2}-3x^{2})(1-y^{2})^{\frac{1}{2}}-\frac{3}{4}y\arcsin
y,...\}
\end{align*}
and%
\begin{align*}
&  \{0,\quad(1-y^{2}),\quad0,\quad2x(1-y^{2}),\quad-2y(1-y^{2})^{\frac{1}{2}%
},\\
&  (3x^{2}-\frac{3}{4}y^{2})(1-y^{2})-\frac{9}{4}y(1-y^{2})^{\frac{1}{2}%
}\arcsin y,\quad-6xy(1-y^{2})^{\frac{1}{2}},...\}.
\end{align*}
Thus, we obtain the following complete system of solutions of (\ref{Beltrami})
with the proportionality factor $\alpha$ defined by (\ref{alphaexample}),%
\[
\overrightarrow{B}_{0}=\left(
\begin{array}
[c]{c}%
y\\
0\\
(1-y^{2})^{\frac{1}{2}}%
\end{array}
\right)  ,\qquad\overrightarrow{B}_{1}=\left(
\begin{array}
[c]{c}%
xy\\
(1-y^{2})\\
x(1-y^{2})^{\frac{1}{2}}%
\end{array}
\right)  ,\qquad\overrightarrow{B}_{2}=\left(
\begin{array}
[c]{c}%
(1-y^{2})^{\frac{1}{2}}\\
0\\
-y
\end{array}
\right)  ,
\]%
\[
\overrightarrow{B}_{3}=\left(
\begin{array}
[c]{c}%
\frac{3}{4}(1-y^{2})^{\frac{1}{2}}\arcsin y+y(x^{2}-\frac{3}{4}y^{2}+\frac
{5}{4})\\
2x(1-y^{2})\\
(1-y^{2})^{\frac{1}{2}}\left(  x^{2}-\frac{1}{4}y^{2}-\frac{3}{4}%
\frac{y\arcsin y}{(1-y^{2})^{\frac{1}{2}}}\right)
\end{array}
\right)  ,\qquad\overrightarrow{B}_{4}=\left(
\begin{array}
[c]{c}%
2x(1-y^{2})^{\frac{1}{2}}\\
-2y(1-y^{2})^{\frac{1}{2}}\\
-2xy
\end{array}
\right)  ,
\]%
\[
\overrightarrow{B}_{5}=\left(
\begin{array}
[c]{c}%
\frac{9}{4}x(1-y^{2})^{\frac{1}{2}}\arcsin y+y\left(  x^{3}-\frac{9}{4}%
xy^{2}+\frac{15}{4}x\right)  \\
(3x^{2}-\frac{3}{4}y^{2})(1-y^{2})-\frac{9}{4}y(1-y^{2})^{\frac{1}{2}}\arcsin
y\\
(1-y^{2})^{\frac{1}{2}}\left(  x^{3}-3x\left(  \frac{1}{4}y^{2}+\frac{3}%
{4}\frac{y\arcsin y}{(1-y^{2})^{\frac{1}{2}}}\right)  \right)
\end{array}
\right)  ,
\]%
\[
\overrightarrow{B}_{6}=\left(
\begin{array}
[c]{c}%
-(\frac{9}{4}y^{2}-3x^{2})(1-y^{2})^{\frac{1}{2}}-\frac{3}{4}y\arcsin y\\
-6xy(1-y^{2})^{\frac{1}{2}}\\
-3x^{2}y+\frac{3}{4}y(1+y^{2})-\frac{3}{4}(1-y^{2})^{\frac{1}{2}}\arcsin y
\end{array}
\right)  ,
\]%
\[
\ldots.
\]

\end{example}

\section{Concluding remarks}

We presented two new results.

\begin{enumerate}
\item We showed that the system of equations describing Beltrami fields on the
plane can be reduced to a Vekua equation of the form (\ref{Vekuamain})
whenever any particular solution of the corresponding second-order equation of
the form (\ref{maineqalpha}) or (\ref{Schrodalpha}) is known.

\item In the case when the proportionality factor $\alpha$ depends on one
Cartesian variable we obtain a particular solution explicitly, construct the
corresponding Vekua equation (section 4), and solve it in the sense that a
complete system of solutions is obtained which gives us a complete system of
solutions of the original vector equation describing Beltrami fields.
\end{enumerate}

Of course not always the integrals involved in the construction of the
complete system of solutions are sufficiently easy to evaluate explicitly as
in the example \ref{ExBeltrami}. Nevertheless our numerical experiments
confirm that in general the formal powers and hence the solutions of
(\ref{Beltrami}) can be calculated with a remarkable accuracy. For example,
the vector $\overrightarrow{B}_{40}$ (see notations in the example
\ref{ExBeltrami}) in the Matlab 7 package on a usual PC can be calculated with
a precision of the order 10$^{-4}$. Thus, the use of formal powers for
numerical solution of boundary value problems corresponding to (\ref{Beltrami}%
) and more generally to equations of the form (\ref{maineq}) is really
promising. The work in this direction will be reported elsewhere.

\end{document}